\documentclass{cjaa}                    % preprint, the final version

\input{psfig.sty}                    %for PS/EPS graphics inclusion

\def\ergs{\rm \  \ erg \ s^{-1}} 

\def\msun{\rm M_{\odot}}
\begin{document}

   \title{The Structure and Variability of Sagittarius A*: Zooming in to
	the Supermassive Black Hole at the Galactic Center }

   \author{Feng Yuan
      \inst{1}\mailto{fyuan@mpifr-bonn.mpg.de}
%% Please move "\mailto{}" to the corresponding author of the paper
%% For single author or all the authors from an institute, use "\inst{}"
   \and Jun-Hui Zhao
      \inst{2}
      }

   \institute{Max-Planck-Institut f\"{u}r Radioastronomie,
          Auf dem H\"{u}gel 69, D-53121 Bonn, Germany\\
   \email{fyuan@mpifr-bonn.mpg.de}
       \and
       Harvard-Smithsonian Center for Astrophysics, 60 
       Garden Street, MS 78, Cambridge, MA 02138, USA\\
          }

   \date{Received~~2002~~~~~~~~~~~~~~~ ; accepted~~2002~~~~~~~~~~~~~~ }

\authorrunning{F. Yuan \& J.-H. Zhao}            %author_head in even pages
\titlerunning{The Structure and Variability of Sagittarius A$^*$}
     
\abstract{
The Galactic center provides a unique astrophysical
laboratory for us to study various astrophysical processes.
In this paper, we review and outline the latest results
from observations of Sgr~A$^*$ in terms of source structure 
and variations in flux density. Sgr~A$^*$ phenomenon represents 
a typical case of low radiative efficiency accretion flow 
surrounding a supermassive black hole in low luminosity AGNs. 
Many pending astrophysical problems found from observations 
of Sgr A$^*$ have challenged the existing astrophysical theories. 
Current theoretical models of Sgr A$^*$ are also reviewed.
   }

\maketitle
%
%________________________________________________ sections below
%
\section{Introduction}           %% first-level section will be auto-capitalized

Sgr A$^*$ is an extremely compact and low luminosity source and
is believed to be associated with the supermassive black hole
at the Galactic center. The nature of its compactness and low
radiative luminosity has been puzzling since the discovery of this
intriguing  radio compact source at the center of the Galaxy
(Balick \& Brown 1974). Recent observations at the wavelengths
from radio to X-ray  ({\it e.g.}, a recent review by Melia \& Falcke 2001)
have shown growing evidence for its association with a 
2.6$\times10^6$ M$_\odot$ super massive
black hole at the dynamic center of the Galaxy ({\it e.g.}, 
Ekart \& Genzel 1997; Menten {\it et al.} 1997; Ghez {\it et al.} 1998; 
Backer \& Sramek 1999; Reid {\it et al.} 1999).
The Sgr~A* phenomenon has provided substantial details in understanding
the physics associated with a low radiative efficiency accretion
flow around a supermassive black hole. In the past decade,
numerous models have been proposed for Sgr A$^*$ ({\it e.g.} Melia 1992;
 Falcke {\it et al.} 1993; Narayan {\it et al.} 1998;
Yuan {\it et al.} 2002). An advection-dominated accretion flow (ADAF)
suggested by Narayan {\it et al.} (1998) appears to have
successfully provided
a dynamical model that can sustain a low radiative efficiency accretion
flow around a supermassive black hole. However, recent observations  
at radio and
X-ray suggests
that the ADAF model appears to be not good enough for Sgr A$^*$.
The nature of Sgr A$^*$ is still far from clear. In this paper,
we will go through the latest results from recent observations
in the aspects of spectrum, source structure and time variability
in flux densities. Models for Sgr A$^*$ are also discussed.  

%% ChJAA editors DO NOT use \cite{} for citation, \ref and \label for
%% cross-references of Table/Figure in publication version, so ChJAA prefers you giving
%% the year as Michel et al. 1992, and writting Table~1 or Fig.~1 and so forth.
%% However, that will make authors inconvenient in adjusting/adding/removing
%% text/table/figure.

\section{Results from Observations}
%\label{sect:Obs}

\subsection{Spectrum}

Because of its proximity and uniqueness, Sgr A$^*$ is a primary target
that has been extensively observed in the past three decades.
 Fig. 1 shows
the well-known spectrum of Sgr A$^*$, indicating that the
radiation flux density from this source peaks at $\sim$1000 GHz
(or 0.3 mm).
The radio spectrum seems to be composed of two components, with a break
frequency at $\sim$50 GHz (e.g., Wright \& Backer 1993; 
Morris \& Serabyn 1996; Falcke {\it et. al.} 1998; Falcke 1999). 
Below this frequency, the spectrum can be characterized 
by a $ \alpha=0.3$ power-law ($S_{\nu}\propto \nu^{\alpha}$), while
above the break frequency, the spectrum presents an excess which peaked 
at $10^{12}$Hz. This components has been referred as 
a sub-millimeter bump (hereafter sub-mm bump). The spectral index of this component
appears to be $ \alpha>0.5$ during a flare state and 
$ \alpha\sim0.25$ in a quiescent state.
  At higher frequencies above the sub-mm bump,
the spectrum drops off steeply as indicated by the infrared observations
(Rieke \& Lebofsky 1982). Most of the infrared data only give an
upper limit in detection
(Telesco, Davidson, \& Werner 1996;  Cotera et al. 1999;
Genzel \& Eckart 1999). Sgr A* may have been detected at 2.2 $\mu$m
in one observation (Genzel {\it et al.} 1997)
but this is the only possible detection to date. 
The source has not been detected
 in the optical/UV bands
due to the strong extinction (visual extinction of 30 mag.)
towards the Galactic center. In general, we believe that 
the emission
from the optical and UV bands is unlikely to exceed the peak at 
sub-millimeters.

\begin{figure}
\psfig{file=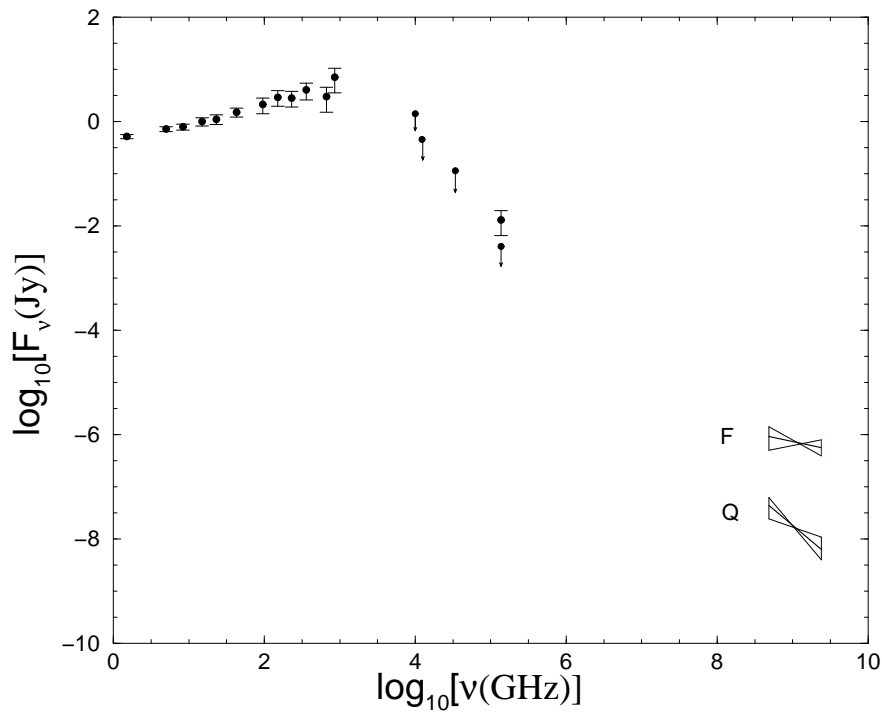,width=0.8\textwidth,angle=270}
\caption{A spectrum of Sgr A$^*$.
The radio and IR data are compiled by Melia \& Falcke (2001). 
The short solid
lines in the X-ray error boxes show the best fit
to the {\em Chandra} observation by a power-law
model in Baganoff et al. (2001).}
\end{figure}

The X-ray spectrum 
shown in Fig. 1 is derived from the Chandra 
observations (Baganoff et al. 2001, 2002) which
is believed to be the first firm detection in X-ray band. 
The two sets of data show  two states, namely quiescent and 
flare states, denoted by
``Q'' and ``F'' respectively. The luminosity and photon index are
$L_x=2.2 \times 10^{33}\ergs$ and $\Gamma\sim 1.5-2.7$
for the quiescent state, and $L_x=10^{35}\ergs$ and $\Gamma\sim 0.7-1.8$
for the flare state. 

The bolometric luminosity is inferred to be $L \sim 10^{-8.5}L_{\rm Edd}$
if the mass of the black hole is $M=2.6\times 10^6\msun$. 
Sgr A$^*$ is an extremely dim AGN. 
The spectrum of Sgr A$^*$ is quite different from those 
observed in the 
luminous AGNs where
the luminosity usually peak at optical/UV bands (so-called Big-Blue-Bump). 
The Sgr A$^*$ type of spectrum appears to be common
in low luminosity AGNs (LLAGNs; {\it e.g.} Ho 1999). 
The difference in the spectra between luminous and low luminous
AGNs indicates 
that a powering mechanism or engine
operating in the center of LLAGNs may  indeed differ from
that in the luminous AGNs.  Sgr A$^*$ is an excellent case
for us to study the astrophysical processes occurring in
the centers of  low luminosity
 AGNs. 

\subsection{Structure}

It has been proved difficult in determining the intrinsic structure
of Sgr A$^*$ because of its compactness and the scattering in the interstellar
medium (ISM) at radio wavelengths ({\it e.g.} Davies, Walsh, \& Booth 1976).
The scattering of the ISM results in a $\lambda^2$ dependence of its
diameter as a function of the observed wavelengths.

The apparent images of the source show elongated structure roughly
in EW with a constant ratio of 2 between major-to-minor axes
({\it e.g.} Lo {\it et al.} 1998). The elongation of the apparent source structure is
thought to be
due to the anisotropy of the magnetized ISM.
Attempts and 
efforts have been made for a decade in determining the intrinsic structure
of Sgr A$^*$ using VLBI technique at millimeter wavelengths
(Backer {\it et al.} 1993;
 Krichbaum {\it et al.} 1993; Rogers {\it et al.} 1994;
Bower\& Backer 1998, Lo {\it et al.} 1998;
Doeleman {\it et al.}, 2001).
Fig. 2 illustrates  the apparent sizes along both the major and minor
axes versus wavelength.  The illustration shows
 that up to the wavelength 7 mm, the apparent
image along the major axis is clearly dominated by the scattering.
At 7 mm, along the minor axis,  a deduced $0.20\pm0.06$ 
mas deviation from
the scattering
size may  suggest that
an intrinsic structure  begins to be revealed in NS.
At millimeter wavelengths, the scattering effect
becomes relatively less severe.  The latest results from 
observations with a
VLBI network at 3 mm suggest that the position angle of
the major axis tends to differ from the constant value of
80 degree observed at the longer wavelengths.
The apparent source structure
($\theta_{maj}=0.34\pm0.14$ mas, $\theta_{min}=0.17\pm0.12$,
P.A.$=22\pm20$ deg)
derived from an elliptical model fitting to the surface brightness
distribution appears to deviate from the scattering
structure  that is extrapolated from the longer wavelengths
with a $\lambda^2$-dependence, indicating the intrinsic
extension of the source may be indeed  in NS.
However, the best-fit circular
Gaussian model of FWHM 0.18$\pm0.02$ mas to the data suggests
that the apparent source structure may be still comparable to the
scattering size. The difficulty in determination
of the source structure at 3 mm is due to the limited UV coverage in NS
and the large atmosphere attenuation which results an uncertainty
in calibrating the visibility data. Nevertheless, 
the intrinsic source size at 3 mm is constrained to be less than 0.27 mas
(Doeleman {\it et al.}, 2001). This new result appears to be consistent with
the wavelength dependence of the intrinsic size of Sgr A$^*$ as 
derived from the 7 mm and 1.4 mm measurements:
     $$ \theta_{int} = 0.08 \lambda_{mm}^{0.9} ~~ mas. $$

\begin{figure}
\vspace{0cm}
\psfig{file=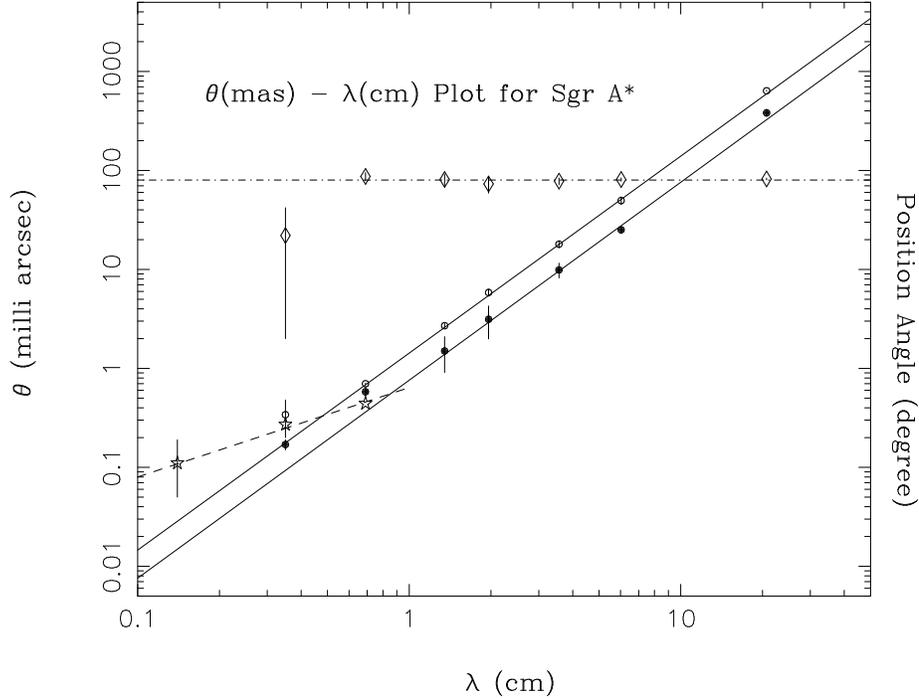,width=0.8\textwidth,angle=270}
\vspace{-4cm}
\caption{The source structure of Sgr A$^*$.
The measurements at 6, 3.6, 2, 1.3 and 0.7 cm are taken
from Lo {\it et. al} 1998. The measurements at 20 cm is from
Yusef-Zadeh {\it et. al} 1994. The 3 mm measurements use
Doeleman {\it et. al.} 2001. The measurement at 1.4 mm 
is from Krichbaum {\it et. al.} 1999. The solid lines are
the scattering size fitting to the observed sizes along both
the major axis (open circles) and the minor axis (solid dots).
The open diamonds indicate the position angle of the apparent 
elongation. The dashed line is a power law fit to the intrinsic
sizes  derived from the measurements at 7, 3 and 1.4 mm (open stars).
}
\end{figure}

The physical size corresponding to 0.27 mas for a black hole of 2.6
$\times10^6$ M$_\odot$ is about 40 R$_{\rm s}$, where 
R$_{\rm s}$ = 7.7 $\times$ 10$^{11}$ cm is the Schwarzschild radius
assuming the black hole mass to be 2.6 $\times10^6$ M$_\odot$.
The radio plasma appears to be highly confined to a small
volume around the super massive black hole.

\subsection{ Flux Density Variation }

The variations in radio flux density observed from Sgr A$^*$
are established since the discovery of this intriguing
radio compact source at the Galactic center in 1974 (Brown \& Lo 1982;
and Zhao {\it et al.} 1989). The nature of the radio variability
has not been well understood.

Based on observations of flux density variation at 1.3 and
0.8 mm with JCMT,
Gwinn {\it et al.} (1991) found that there were no significant
variations in a time scale of 1 sec to a day while they were searching
for reflective scintillation.

During the period of 1990-1993, a regular VLA
flux density monitoring program was carried out suggesting
that the amplitude variation increased towards short wavelengths
and that the rate of radio flares appeared to be about three per year
(Zhao {\it et al.} 1992; and Zhao \& Goss 1993).
The typical time scale of these radio flares is about
a month.  Large-amplitude
fluctuations in the flux density observed at 3mm (Wright \& Backer 1993;
Tsuboi, Miyazaki and Tsutsumi 1999) appeared to be
consistent with what were observed at centimeter wavelengths.
Based on the radio-monitoring data obtained with the 3.5 km Green Bank
Interferometer (GBI) at 11 and 3.6 cm, a characteristic time scale
of 50-200 days was observed at both wavelengths and  the
structure function of 11 cm data suggested a quasi-periodic variation
with a period of 57 days (Falcke 1999).

A presence of a 106 day cycle in the radio variability of Sgr A$^*$
was revealed based on an analysis of data observed with the VLA over
two decades (Zhao, Bower and Goss 2001). The results derived from
VLA archived data show
that the pulsed components with a spectral index of 1.0$\pm$0.1
and an amplitude $\Delta S=0.42\pm0.04$ Jy and a characteristic timescale
of $\Delta t_{FWHM}\approx 25\pm 5$ days. The lack of a VLBI detection of
secondary component suggests that the variability arises from Sgr A$^*$
on a scale of $\sim$ 5 AU.

A new regular VLA monitoring program at wavelengths 2, 1.3 and 0.7 cm
with a typical sampling
interval of a week has been launched since the mid of year 2000.
The preliminary results based on the data obtained in the past 500
days show that the fluctuation in flux density appears to  persist
(McGary {\it et al.} 2002; Bower {\it et al.} 2002).
The power spectral density (psd) profiles show that the cycle lengths
of the pulsed signals are 133$\pm$3, 135$\pm$3 and 121$\pm$3 days
for observing wavelengths of 2, 1.3 and 0.7 cm, respectively.
The length of the cycle time appears to become
longer as compared to the 106 day cycle derived from the
VLA archive data in the period between 1977 to 1999.
In addition, the lags derived from the light curves suggest
that short wavelengths tend to peak first.

\begin{figure}
\vspace{0.5cm}
\psfig{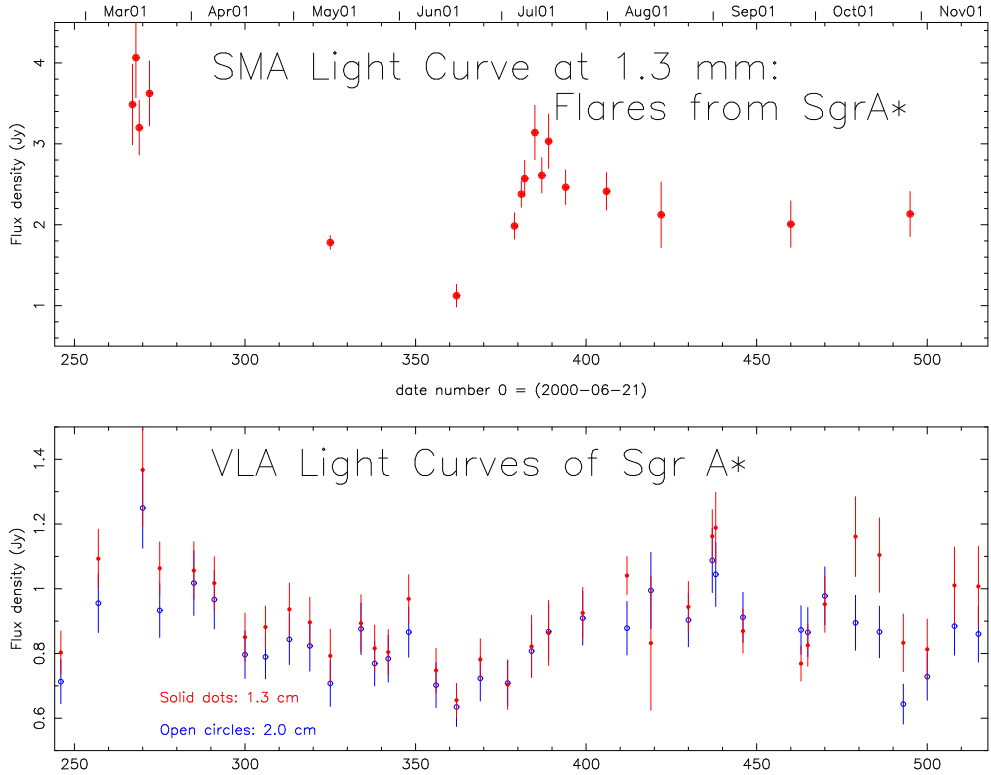}
\vspace{0.5cm}
\caption{The light curve of Sgr A$^*$ observed at 1.3 mm with the SMA of
Smithsonian Astrophysical Observatory (upper panel) is compared
with the radio light curves (lower panel) 
observed at 1.3 and 2 cm using the
VLA of NRAO in the period between March to November year
2001 (Zhao {\it et al.} 2002; McGary {\it et al.} 2002). The preliminary results of the 1.3 mm observations
with the partially completed SMA were reported in 
the 199 AAS meeting in Washington D.C. (Young {\it et al.} 2002
and Zhao {\it et al.} 2002).
}
\end{figure}

A short millimeter and sub-millimeter monitoring
program has also been carried out with the Sub-millimeter Array (SMA)
of Smithsonian Astrophysical Observatory since mid March of 2001
(Zhao {\it et al.} 2002).
A total of 19 epochs of observations were made at 1.3 and 0.87 mm.
The SMA light curve of Sgr A$^*$ at 1.3 mm suggests that two flares
from Sgr A$^*$. The March flare (started from 4.1$\pm$0.5 Jy after an unseen
peak and decreased to 1.1$\pm$0.15 in three months) appeared to be relatively
stronger than that in July (3.1$\pm$0.3Jy). The two flares
show a good correlation between the 1 mm flux densities and those measured  
with VLA at 1 to 2 cm. Considering the fact that the peak
was not seen, there might be a time-delay between 1mm and 1 cm during the
March event. During the second event, the VLA light curves show a slow increase
in flux density suggesting a significant delay. In addition,
the spectral index varied from 0.5$\pm$0.1 of the March flare to 0.23$\pm$0.07
in the June minimum and then increased to 0.6$\pm$0.1 in the July flare.
The steep rising spectra of Sgr A$^*$ during the flares do show
an excess of short-mm/sub-mm but this sub-mm bump does not appear to be a
stable component. In addition, time separation between
the two flares (March and July) is $\ge$ 120 days, which is consistent
with the cycle length derived from PSD analysis of the new VLA
data. We note that the observations at 1.3mm were poorly sampled
but the weekly VLA observations show there were no additional
significant flares between the March and July events.

An independent monitoring program at 3 mm was
carried out with  Nobeyama Millimeter Array (NMA) since 1996
and several flares were observed in the period of 1996 to 2001
(Tsuboi, Miyazaki and Tsutsumi 1999; Miyazaki, Tsutsumi and Tsuboi
2001). Due to the fact that Sgr A$^*$ monitoring observations
require long baselines or large array configurations in order
to separate the point source of Sgr A$^*$ from the
surrounding extended HII emission, the monitoring observations with NMA
were only available in a period from late fall to early
spring each year.  The NMA light curve
produced from the nearly 60 measurements at 90 and 102 GHz
shows several flares during the course of monitoring.
There are several gaps in the NMA sampling. 
 A profile produced
by folding the 3mm data into a 106 day module also shows two distinguishable
phases: one with flares and another representing quiescent states
(Tsuboi {\it et al.} 2002, private communication). It will be interesting 
to see how the absolute
phase at 3 mm is related to the phases of the flux density fluctuation 
observed at other wavelengths.

Finally, the flare of Sgr A$^*$ observed with Chandra
in the X-ray on 27 October 2000 (Baganoff {\it et al.} 2001) appeared to be followed by radio
peaks about a week later
at all three VLA wavelengths (2, 1.3 and 0.7 cm; McGary {\it et al.} 2002).
The X-ray flare is strong (a  factor of 50 increase)
and short (timescale of 1 hr) and the radio event is relatively weak
($\sim$30\% increase) and long (timescale of a few weeks).
The correlation between the X-ray  and radio flares are consistent with
a picture that the X-ray flares may be related to
mass ejection originating from active regions in the inner
part of the accretion disk. The high-energy electrons
accelerated through a process such as shock or magnetic reconnection 
in the active region are transported
to the outer region via a bulk motion of an outflow (or a jet).
At a certain distance from the
super massive black hole, the outflow tends to be disrupted.
The high-energy particles along with the disrupted outflow
lose their outwards bulk momenta and 
are then recycled back to the disk. 
Assembling these processes may produce a variation cycle in flux density
if an MHD convection is dominant in the dynamics of the accretion fluid.
The observed quasi-periodic
cycle in flux density fluctuation may have provided evidence for
that a convection driven by an MHD dynamo may indeed operate
in the accretion disk at Galactic center. 
The discussion here provides a qualitative scenario in understanding
the observations.
How to cope with the observed
variability  is still a challenging problem
 to the current astrophysical theories in
modeling the low radiative efficiency accretion flow
({\it e. g.}, Igumenshchev \& Narayan 2002; Narayan 2002).

\section{Theoretical Models}

Almost all models proposed 
for Sgr A$^*$ are based on accretion onto the
central massive black hole. In general, there are three 
types of models for Sgr A$^*$, namely accretion, 
non-accretion flow (such as jet or outflow wind) and their 
combination. The accretion rate derived from the Chandra observations
(Baganoff {\it et al.} 2002) is about 10$^{-4}$ \.M$_{Edd}$,
where \.M$_{Edd}$ is the Eddington accretion rate. This rate 
would
result an unreasonably high
luminosity if  the radiative efficiency is assumed to 
be the canonical value of 
a standard thin disk ($\eta\sim0.1$). 
Then, all the models must confront with the
low radiative efficiency problem.  

\subsection{Accretion Flow Models}

The first accretion model proposed is spherical accretion model (Melia 1992; 1994;
Melia, Liu, \& Coker 2001). In the latest version of this model, 
considering that 
 the specific angular momentum of accretion flow is likely very low,
the accretion flow is assumed to be
free-fall until a Keplerian disk is formed within a small
``circularization'' radius, $r \sim 5-10 {\rm R_s}$. The electrons in the
small Keplerian disk can attain a very high temperature through some
magnetic heating processes, $T_e \ga 10^{11}$K, a significant mass loss
is assumed so that the density of the particles is extremely low, 
$n\sim 10^{6-7}$ electrons cm$^{-3}$,
and the magnetic field is about 10-20G. Thus the low efficiency
is achieved due to the mass loss in this model. 
The synchrotron and self-Compton
emission from the small Keplerian disk 
can be utilized to explain the observed
sub-mm bump and the X-ray spectrum, respectively. 

A transition from a spherical flow to a small disk is interesting
but  the formation of the small Keplerian disk
may not be a necessary result of 
low angular momentum accretion. An accretion flow with very low
angular momentum can still be described by a ``disk'', although such
accretion may belong to the different modes (Bondi-like type or disk-like
type), as shown by Yuan (1999) (see also Abramowicz \& Zurek 1981;
Abramowicz 1998). 
Considering their spectral prediction, this model can't
fit the low-frequency radio spectrum below the sub-mm bump. In fact, 
as shown by Liu \& Melia (2001), 
the thermal synchrotron emission from an accretion flow
is unlikely to fit the low-frequency radio spectrum.
In the X-ray band, this model suggests that
almost all the flux arises from an extremely compact region via
the inverse Compton scattering
of synchrotron emission. 
However, the Chandra X-ray observation suggests
that the source is extended  in its quiescent state
(Baganoff et al. 2002). 
%The inverse Compton scattering may be possible
%to explain the X-ray flare observed by Baganoff {\it et al.} (2001).

Another accretion flow model for Sgr A$^*$ is ADAF (Narayan et al. 1995; 
Manmoto et al. 1997; Narayan et al. 1998).  The most attractive feature of
the ADAF model is its ability to
explain the unusual low-luminosity of Sgr A$^*$ given the
relatively abundant accretion material.
The basic assumption  in this model is that 
the viscous dissipation prefers to 
heating ions only and the Coulomb collision is
the only coupling mechanism between 
ions and electrons. Consequently, the dissipation
energy due to the viscosity is stored in the ions and advected into
the black hole rather than radiated away (Ichimaru 1977; Rees et
al. 1982; Narayan \& Yi 1994, 1995; Abramowicz et al. 1995).
In the innermost region of ADAF the particle
density is about $10^{8-9}$ cm$^{-3}$, $T_e\sim 10^{10}$K, 
which is quite different
from those used in the small disk (Melia {\it et al.} 2001).

In the application of ADAF to Sgr A$^*$, the radio spectrum is produced
by the thermal synchrotron emission in the innermost region of the disk.
The X-rays are mainly due to bremsstrahlung radiation of the thermal
electrons in a large range of radii $\sim 10^4-10^5 R_{\rm s}$, which is
in an excellent agreement with the observed extension  of the source.
However, the observed radio flux density appears to be a problem
to the ADAF. By fitting a spectrum produced from
ADAF to both the X-ray flux density in the
Q state and the sub-millimeter peak, 
the radio flux density derived from ADAF alone is 
an order of magnitude below the radio flux density 
of Sgr A$^*$ observed at centimeter wavelengths. 

In the X-ray band, the spectrum predicted by a canonical 
ADAF model, in which most of the dissipation energy is 
used to heat ions mainly, is somewhat too hard 
(but see Quataert 2002 for an explanation). In addition,
the canonical ADAF model appears to be
difficult in explaining  the possible rapid fluctuation detected
in the Q state by Baganoff et al. (2002) (Yuan, Markoff, \& Falcke 2002).
However, the two problems can be solved if the electrons
are heated directly by a
moderately large fraction of the viscous dissipation
energy.  In such a case,  the second order
Comptonization of the synchrotron emission will contribute
significantly to the X-ray spectrum (see Figure 4. of Narayan 2002). 
Taking into account strong winds from ADAF can't make the fitting better
if the wind is assumed not to  radiate (Yuan, Markoff, \& Falcke 2002).
Thus a second component with a soft spectrum and a rapidly variable 
nature
may be needed to explain the observed X-ray emission.

There are other low radiative efficiency accretion models which can 
potentially explain the observations of Sgr A$^*$ such as 
convection-dominated accretion flows (CDAFs). Recent
3-D MHD simulations of spherical accretion conducted by 
Igumenshchev and Narayan (2002) suggests 
that the MHD accretion flow will eventually transit to a 
state of self-sustained convection via the MHD buoyancy-induced motions,
magnetic field reconnection and gas heating. 
An MHD CDAF coupling with relevant high energy particle
cooling mechanism will be interesting in understanding 
the Sgr A$^*$ phenomenon. 
A recent review by Narayan (2002) covers the theoretical
front of the CDAF in details. 

\subsection{Jet Model}

In the second kind of model for Sgr A$^*$, it is assumed that the accretion
flow contribute little to the total luminosity of Sgr A$^*$ and the flux
comes from a jet/outflow (Reynolds \& McKee 1980; 
Falcke, Mannheim, \& Biermann 1993; Falcke \& Biermann 1999; Falcke 
\& Markoff 2000). This is based on the idea of Blandford \& K\"onigl (1979) 
that flat-spectrum radio cores are best explained by 
synchrotron-emitting jets. In the most updated version of jet model 
for Sgr A$^*$, the sub-mm bump is due to the synchrotron emission from the
acceleration region in the base of the jet, called nozzle. The parameters
describing the nozzle is $n\sim 10^{6-7}$ cm$^{-3}$, $T_e\sim 10^{11}$K, and
$B \sim 20$G. The parameters used here 
are similar to  those in the models proposed 
by  Beckert \& Dushcl (1997) and 
Melia, Liu, \& Coker (2001).

Following the
evolution of the plasma in the nozzle described by the above 
parameters on its way out using the Euler equation, the emission from the
jet can well interpret the low-frequency radio emission of Sgr A$^*$.
As mentioned above, the radio spectrum 
is difficult to be explained 
by the either Bondi-accretion flow or ADAF. In
addition, the inverse Compton emission of the plasma in the nozzle 
can explain the soft X-ray spectrum and its possible rapid variability.

But the remaining important problem in the jet model is 
why the parameters of the jet possess the required values,
particularly in reference to the inferred underlying accretion disk.
Previous ideas of a standard optically thick accretion disk for Sgr
A* do not seem to work because the predicted IR flux
from a standard thin disk with a reasonable
accretion rate would be several orders of magnitude higher than the
observed IR upper limit (Falcke \& Melia 1997).
In addition, the jet model can't explain why an extended
X-ray source is observed since the radius and height of the nozzle is only 
$\sim 4GM/c^2, 10GM/c^2$, respectively. 

\subsection{Jet-ADAF model}

The observations to Sgr A$^*$ strongly suggest that a jet 
and an ADAF may co-exist in the immediate environment
surrounding the super massive black hole at the Galactic center.
The emission of Sgr A$^*$ might be indeed produced from 
both the jet and the ADAF.
On one hand, 
the flat radio spectrum 
is best explained by the jet emission, 
and the possible rapid variability
at the X-ray band and the soft X-ray spectrum 
also suggest that the inverse
Compton emission from the jet nozzle should contribute partly. 
On the 
other hand, the observed extended feature of the source strongly 
indicates the existence of 
the bremsstrahlung emission from ADAF. Therefore, it 
is crucial to consider the jet and accretion flow as a coupled system
in Sgr A$^*$. We need also to consider what are their respective roles if both
are truly present in Sgr A$^*$.

Yuan, Markoff \& Falcke (2002) propose a coupled jet-ADAF model
to describe the  
emission of Sgr A$^*$.
This model can also provide a convincing explanation to
the spectrum during the X-ray 
flare state of Sgr A$^*$ ({\it e.g.}~Markoff et al. 2001). This is a good 
advantage using the jet-ADAF coupling model
in explaining observed spectrum and variability
as compared to the models using accretion flow alone.
Within r$\sim 4GM/c^2$, a fraction 
of the accretion plasma is ejected out of the ADAF and
forms the jet. A standing shock occurs in this process
since the accretion flow is
radially supersonic at r$\sim4GM/c^2$. From the shock transition condition,
we can determine the post-shock temperature. But 
the post-shock density depends on
what a fraction of the 
 accretion flow will be transfered into the jet which
is related to the jet-formation physics and is hard 
to determine due to the fact the jet-formation process
is still poorly understood. 
We use  parameters to describe such a process. As usual, the 
magnetic field is determined
by a parameter describing the ratio between the magnetic energy to the
post-shock thermal energy. Once the parameters
describing the nozzle are determined, the emission from the nozzle and jet 
can be calculated. The resulting spectrum
from Sgr A$^*$ is the sum of the jet and ADAF. 
The X-ray emission is partly due to the bremsstrahlung emission from
the ADAF and partly due to the inverse Compton from the nozzle, thus
the source is extended and the rapid variability can be explained.
The low-frequency radio spectrum is almost completely dominated by the 
emission from the jet while the sub-mm bump is the sum of the emissions 
from the jet and ADAF. 

Although the jet-ADAF model seems to be able to
explain the spectrum of 
Sgr A$^*$ ranging from radio to X-ray very well,
a lot of work still need to be done.
The most important part of the uncompleted work is to fill in
the details for 
the coupling between the jet and disk. We are still not certain
how the accretion flow is ejected out
of the disk and how jet material is accelerated via the nozzle. 

\section{Summary}

In this paper we review the observations of Sgr A$^*$ in the aspects
of its structure and
variation in flux density and outline the current theoretical
models of this source. 
Most of these models 
focus only on the fitting to the observed spectrum.
The structure
of Sgr A$^*$ is discussed in the jet model by Falcke \& Markoff (2000).
However, based on the VLBI observations, the reality is that
down to an angular scale of 0.27 mas ($\sim$40 Schwarzschild radii)
there is still no convincing evidence for intrinsic 
extended structure.
The variation in flux density at 1cm to 1mm shows that
a flux density varies by 50\% at 1 cm and by a factor of 3
at 1 mm in a typical time scale of 1 month. 
The variation in flux density between a flare state and a quiescent state
suggests a fluctuation cycle of four months. 
If the the variation in flux
density is related to a jet or outflow, a
fluctuation in intrinsic source size around 0.2 mas
is expected at the short wavelengths. Due to the strong 
ISM scattering at long wavelengths and availability
in the millimeter VLBI technique, the best observing 
wavelength is at 3 mm. 

Monitoring the flux density at multi-wavelengths
between centimeter, millimeters, sub-millimeter, IR and X-ray
is necessary in order to understanding the radiative
cooling in the high energy processes occurring in this
source. 

The fluctuation cycle observed at the wavelengths
between  short centimeters
to short millimeters is interesting but has not been
well understood yet.  Understanding the fluctuation cycle
may naturally lead us to understanding the dynamical link between
the accretion inflow and jet outflow that may co-exist in Sgr A$^*$.
In other words, the variability will put a strong constraint
on the theoretical models of Sgr A$^*$

Moreover,
our understanding to Sgr A$^*$ phenomena will help us in understanding 
the puzzling nature of LLAGNs.
                
\begin{acknowledgements}
We thank the organizers of this conference for their invitation 
and hospitality. 
\end{acknowledgements}

%\appendix                  %%appendicial material is supported

%\section{What is SCI?}
% SCI is the abbreviation of Science Citation Index system powered by
% the Institute for Scientific Information (ISI).
% For details please visit http://www.isinet.com/isi/journals/index.html

%\section{This shows the use of appendix}
% When you compress a *.ps file with gzip.exe in DOS/Windows, you get *.psz.
% Linux's counterpart of the DOS/Windows pkzip/pkunzip or winzip are zip/unzip.

\end{document}